\documentclass[12pt]{article}
\usepackage{amsfonts}

\usepackage{myart}

\normalbaselines
\input tcilatex.tex

\begin{document}

\author{Yuri A. Rylov}
\title{Model conception of quantum phenomena: logical structure and investigation
methods.}
\date{Institute for Problems in Mechanics, Russian Academy of Sciences\\
101, bld. 1, Vernadski Ave. Moscow, 119526, Russia \\
email: rylov@ipmnet.ru\\
Web site: {$http://rsfq1.physics.sunysb.edu/\symbol{126}rylov/yrylov.htm$}\\
}
\maketitle

\begin{abstract}
One can construct the model conception of quantum phenomena (MCQP) which
relates to the axiomatic conception of quantum phenomena (ACQP), (i.e. to
the conventional quantum mechanics) in the same way, as the statistical
physics relates to thermodynamics. Such a possibility is based on a new
conception of geometry, which admits one to construct such a deterministic
space-time geometry, where motion of free particles is primordially
stochastic. The space-time geometry can be chosen in such a way that
statistical description of random particle motion coincides with the quantum
description. Transition from ACQP\ to MCQP is a result of correction of some
mistakes in the foundation of ACQP. Methods of MCQP in investigation of
quantum phenomena appear to be more subtle and effective, than that of ACQP.
For instance, investigation of the free Dirac equation in framework of MCQP
shows that the Dirac particle is in reality a rotator, i.e. two particles
rotating around their common center of inertia. In the framework of MCQP one
can discover the force field, responsible for pair production, that is
impossible in the framework of ACQP.
\end{abstract}

\newpage

\section{Introduction}

Sometimes investigation of a new class of physical phenomena is carried out
by two stages. At first, the simpler axiomatic conception based on simple
empiric considerations arises. Next, the axiomatic conception is replaced by
the more developed model conception, where axioms of the first stage are
obtained as properties of the model. Theory of thermal phenomena was
developed according to this scheme. At first, the thermodynamics (axiomatic
conception of thermal phenomena, or ACTP) appeared. Next, the statistical
physics (model conception of thermal phenomena, or MCTP) appeared. Axioms of
thermodynamics were obtained as properties of the chaotic molecule motion.

The contemporary quantum theory is the first (axiomatic) stage in the
development of the microcosm physics. Formal evidences of this is an
existence of quantum principles. Appearance of the next (model) stage, where
the quantum principles are consequences of the model, seems to be
unavoidable. The model conception is attractive, because it uses more subtle
and effective mathematical methods of investigation. Besides it gives
boundaries of the axiomatic conception application. We can see this in
example of statistical physics and thermodynamics.

From practical viewpoint, the main difference between the axiomatic and
model stages of a theory lies in mathematical methods of description and
investigation. Methods of axiomatic conception are more rough are rigid.
They can be changed only by a change of axiomatics. This is produced mainly
by introduction of additional suppositions. Mathematical methods of model
conception are more flexible and adequate, because usually the model
parameters are numbers and functions, which can be changed fluently.

Let us imagine, that we do not know methods of statistical physics and try
to investigate nature of crystal anisotropy, using only thermodynamical
methods. Using experimental data we can calculate thermodynamical potentials
and describe macroscopic properties of crystal, but we hardly can calculate
anything theoretically. For description of the next crystal we are forced to
use experimental data again. By means of methods of statistical physics we
can explain and calculate parameters of crystal anisotropy theoretically. At
any rate, methods of model conception appears to be more effective in the
given case.

Something like that we observe in the theory of elementary particles, when
theorists use rough and rigid methods of axiomatic conception (quantum
theory), which do not allow to construct a perfect theory. There is a hope
that mathematical methods of the model conception of quantum phenomena
(MCQP) appear to be more effective, because MCQP does not use the rigid
principles of quantum mechanics. Instead quantum principles MCQP uses
parameters of space-time geometry which can be changed fluently.

We shall refer to the contemporary quantum theory as the axiomatic
conception of quantum phenomena (ACQP). Logical structure of ACQP
distinguishes from the logical structure of MCQP and from that of other well
defined physical conceptions. For instance, we compare the logical structure
ACQP with the logical structure of such a physical conception as the special
relativity theory. The relativity theory appeared in the beginning of the
XXth century as a solution of the problem of large velocities, when the
particle motion with the speed close to the speed of the light cannot be
described correctly in the framework of the classical nonrelativistic
physics. Creation of the special relativity theory was a result of a change
of the fundamental principle of physics. The space-time model has been
changed. The Newtonian space-time model with two independent invariant
distances between two events (the temporal interval and spatial interval)
was replaced by the space-time geometry with one space-time interval between
the two events. Such a solution of the problem of large velocities was
sufficiently fundamental, and the scenario of the special relativity
development was classical in the sense that it reminds the development
scenario of the classical (Newtonian) physics. The last developed in
accordance with the Newton's slogan ''Hypotheses non fingo'' (Not to invent
hypotheses). Logical structure of the special relativity theory and that of
ACQP are shown in figure $1$, where ovals show experimental data and bars
show the statement of the conception.

ACQP arises as a solution of the problem of the atomic spectra discreteness.
This problem appeared in the beginning of the XXth century. As it became
clear later, this problem is a special case of a more general problem of the
small mass particle motion. Experiments show that the particles of small
mass move stochastically, and this stochasticity depends on the mass of the
particle. In the classical (relativistic and nonrelativistic) physics the
motion of a free particle is determined only by the space-time geometry.
This property should be conserved in the modified theory. One should invent
such a space-time geometry, where the motion of a free particle would be
stochastic, and this stochasticity would depend on the particle mass. (By
definition, the geometry is a set of points, or events with a given distance
between any two points). One should note that the geometrical character of
the free motion and a dependence on the particle mass are compatible only in
that case, when the particle mass is a geometrical characteristic of the
particle (the mass geometrization). In the classical (relativistic and
nonrelativistic) physics the particle mass is not geometrized, and the mass
geometrization is a quite new property of a space-time geometry. Space-time
geometries with such unusual properties (stochasticity of free motion and
the mass geometrization) were not known in the beginning of the XXth century
(they were not known until the end of the XXth century). Besides, at that
time the problem of the small mass particles was perceived rather as a
problem of the atomic spectra discreteness.

In the beginning of the XXth century the problem of the small mass particles
cannot be solved on the fundamental level (by a change of the space-time
geometry). The problem was solved on the second (after fundamental) level by
introduction of the quantum principles, which replace some of corollaries of
the true space-time geometry (see figure $1b$). The space-time model remains
to be Newtonian (nonrelativistic), but the logical and mathematical
corollaries of this model which concern the quantum phenomena (motion of the
small mass particles), are blocked. Corollaries of quantum principles are
used instead of them. Such a logical construction, when corollaries of a
false fundamental statement are replaced by the true corollaries of the true
fundamental statement, will be referred to as a compensating conception. The
compensating conception contains simultaneously both statements: the false
fundamental statement and the true corollaries of the true fundamental
statement. The reason, why some true fundamental statement is absent in the
compensating conception may be different. This reason is of no importance,
but it is important that because of the absence of the true fundamental
statement one cannot obtain all its corollaries. One obtains only some of
them, whereas other corollaries of the true fundamental statement remained
to be unknown.

These unknown corollaries appear to be needed, when we consider a new set of
physical phenomena. One can guess right the unknown corollary of the absent
true fundamental statement, provided this corollary is located on the upper
level of the conception statements, i.e. in the case, when this corollary is
close to the experimental data. But in this case the corollary is far from
the true fundamental statement (on the bottom level) and from the quantum
(compensating) principles (on the second level) (see figure $1b$). If we
succeeded to guess right the unknown corollary and to explain some
experimental data, the obtained statement looks as something unrelated with
the quantum principles. This statement is perceived as a new additional
supposition (hypothesis) which is needed for explanation of experimental
data. The more the new experimental data, the more new hypotheses we need.
Thus, we can develop the compensating conception only inventing new
hypotheses, located on the upper level near the experimental data. Discovery
of corollaries, located near the fundamental level, is difficult, because
the true fundamental statement is unknown and one cannot deduce its
corollaries. At the same time it is difficult to guess right these
corollaries, using experimental information, because they connected with the
experimental data by a long logical chains.

The farther from the fundamental level are the new hypotheses located, the
less is the predictability of the physical conception in the relation of the
new experiments. The closer is the new hypotheses to the experimental data,
the easier they can be guessed right. Thus, if the true fundamental
statement is unknown, the development of the physical conception is carried
out in such a way, that the number of additional hypotheses increases,
whereas the predictability of the compensating conception in the relation of
new experiments decreases. To improve the predictability of the compensating
conception, it is necessary to obtain the true fundamental statement. New
experimental data cannot help in this search, because as a rule the logical
chains connecting the true fundamental statement with the experimental data
are too long. One can obtain this fundamental statement only by the logical
way, using the whole set of experimental data.

Studying the history of the quantum theory development in application to
relativistic phenomena and to the theory of elementary particles, one can
discover characteristic evidence of the compensating conception. The number
of additional hypotheses increases, whereas the predictability of the
conception in the relation to new experiments decreases. Besides, there are
additional circumstances conditioned by the long work with the compensating
conception (ACQP). These circumstances impede transition from ACQP to MCQP.
As a rule the contemporary researchers dealing with the relativistic quantum
theory evaluate a new hypothesis as follows. If the new hypothesis explains
some new experimental data, this hypothesis deserves to be accepted and
investigated. If the author of a new hypothesis cannot manifest any
experiment which can be explained by his new hypothesis, such a hypothesis
does not deserve even to be investigated. Such an approach to the test of
new hypotheses is justified only in framework of the compensating
conception, where all new hypotheses are located on the upper level, and
logical chain between the hypothesis and the experimental data is short.
However, this approach to the test of the hypothesis is inadequate, if one
suggests the fundamental statement, located on the bottom level. In this
case the logical chain between the fundamental statement and experimental
data is very long. To explain experimental data and to test this chain, one
needs to develop the investigation methods and mathematical technique,
corresponding to the new fundamental statement. All this is a very difficult
problem, and to test the new hypothesis, which pretends to be a fundamental
statement of the conception, one should use mainly such an argument, as
absence of other additional hypotheses. Another criterion of the fundamental
statement is the fact that it is introduced instead of a false fundamental
statement, but not in addition to it. For instance, if the space-time
geometry is considered to be a fundamental statement of the physical
conception, the transition from ACQP to MCQP is to be accompanied mainly by
a change of the space-time geometry. In this case one should speak rather on
a correction of a mistake in the foundation of the physical conception. Of
course, ultimately an explanation of all experimental data and prediction of
experimental results is the final criterion of the fundamental statement
validity, but this explanation may appear only after a long development of
the mathematical technique, corresponding to the new fundamental statement.

According to MCQP quantum phenomena are a result of stochastic behavior of
microparticles. Statistical description of this stochastic motion leads to a
description of the quantum phenomena. This idea is very old. It was
reasonable after the successful explanation of thermal phenomena as a
chaotic motion of molecules. There were many attempts of this idea
realization, but all these attempts had failed. As a result a sceptic
relation to this idea appeared. Now most of physicists believe that quantum
principles describe correctly the origin of physical phenomena in microcosm,
and the microcosm physics has a quantum nature.

Logical connection between the ACQP and MCQP is shown in figure 2. Problems
of the logical links of MCQP in transition from ACQP to MCQP are marked by
the symbol ''?''. As one can see from this scheme that there are three
obstacles on the way of creation MCQP: (1) inadequate space-time geometry,
which describes the particle motion in microcosm as a deterministic,
although experiments show that this motion is random in reality, (2)
inadequate statistical description, when particles and antiparticles are
considered to be objects of statistical description, whereas objects of
statistical description are to be world lines, which are primary physical
objects in relativistic theory, (3) integration of dynamic equations for
ideal fluid, which is necessary for transformation of wave function and spin
(which are fundamental objects of ACQP) into a method of the ideal fluid
description, i.e. in attributes of a description of physical objects.

Model conception of quantum phenomena (MCQP) could be constructed only after
overcoming of these three obstacles. But each of the said obstacles was a
very difficult problem. Besides, it was necessary to realize that on the way
to creation of MCQP there are these obstacles. To overcome the first
obstacle it was necessary to construct a new conception of geometry
(T-geometry), which contains such a space-time geometry, where the particle
motion be random and the particle mass be geometrized.

The true space-time geometry is such a geometry, where motion of
microparticles. is primordially stochastic, although the geometry in itself
is not random (intervals between the events in such a geometry are
deterministic, but not random). To construct such a geometry (T-geometry),
one needs to go outside the framework of Riemannian geometry, which is the
most general contemporary geometry fitting for the space-time description.
Being flat, uniform and isotropic, the true geometry of the absolute
space-time distinguishes from the Minkowski geometry only in some correction
containing the quantum constant $\hbar $. This correction is essential only
for short space-time intervals, i.e. only in microcosm. Formally,
introduction of this correction is equivalent to introduction of some
fundamental length, but it is a transverse length (thickness of the world
line).

Statistical description of a random motion of particles generated by the
space-time geometry leads to the quantum mechanical description (the quantum
constant $\hbar $ appears in the theory via geometry), in the same way as
the statistical description of chaotic molecule motion leads to
thermodynamics. Essential difference between the two statistical
descriptions lies in the difference between the statistical description for
relativistic and nonrelativistic cases. In the case of the statistical
physics both regular and random components of the particle velocity are
nonrelativistic, whereas in the case of the geometric stochasticity the
random velocity component is always relativistic, although the regular
component may be nonrelativistic. As a result even the nonrelativistic
quantum mechanics appears to be a hidden relativistic theory. There is an
essential difference between the relativistic and nonrelativistic
statistical descriptions. The fact is that the nonrelativistic statistical
description may be probabilistic, i.e. it can be carried out in terms of the
probability density, whereas \textit{the relativistic statistical
description cannot be produced in terms of the probability theory}. The
origin lies in the fact that physical objects to be statistically described
are different in the relativistic and nonrelativistic theories.

In the nonrelativistic theory the physical object is a particle, i.e. a
pointlike (zero-dimensional) object in three-dimensional space, and the
particle world line describes a history of the pointlike object. In other
words, the particle is primary and its world line is secondary. In the 
\textit{consequent} relativistic theory the situation is inverse. The
physical object is the world line, i.e. the one-dimensional line in the
space-time, whereas the particle and the antiparticle are derivative objects
(intersections of the world line with the surface $t=$const). In other
words, in the consequent relativistic theory the world line is primary,
whereas the particle and antiparticle are secondary. The term 'WL' will be
used for the world line considered to be the primary physical object. The
difference in the choice of the primary physical object is conditioned by
different relation to the existence of absolute simultaneity in relativistic
and nonrelativistic physics. Statistical description must be a description
of primary physical objects. Foundation for such a description is the
density of physical objects in the three-dimensional space (for particles)
or in the space-time (for WLs). Density $\rho \left( \mathbf{x}\right) $ of
particles at the point $\mathbf{x}$ is defined by the relation 
\begin{equation}
dN=\rho (\mathbf{x})dV  \label{b1.1}
\end{equation}
where $dN$ is the number of particles in the 3-volume $dV$. The particle
density is defined as a proportionality coefficient between $dN$ and $dV$.
The density $j^{k}(x)$ of WLs is defined by the relation 
\begin{equation}
dN=j^{k}\left( x\right) dS_{k}  \label{a1.2}
\end{equation}
where $dN$ is the flux of WLs through three-dimensional area $dS_{k}$ in the
space-time in vicinity of the point $x$. The quantity $j^{k}(x)$ is the
proportionality coefficient between $dN$ and $dS_{k}$ in vicinity of the
point $x$. The quantity $\rho \left( \mathbf{x}\right) $ is a nonnegative
3-scalar. It can serve as a basis for introduction of the probability
density, whereas the 4-vector $j^{k}(x)$ is not a nonnegative quantity, and
it cannot serve as a basis for introduction of the probability density,
which must be nonnegative quantity.

It is a common practice to think that terms ''probabilistic description''
and ''statistical description'' are synonyms. It is a delusion, because the
probabilistic description is a description, founded on a use of the
probability theory, whereas the statistical description is a description,
dealing with many similar or almost similar objects. Such a set of similar
objects is called statistical ensemble. Statistical description is an
investigation and description of the statistical ensemble properties.
Statistical description without a use of the probability density is
possible. It is necessary only to investigate the statistical ensemble
without a use of the probability theory. We shall consider statistical
ensembles of dynamic or stochastic systems and use essentially the
circumstance, that \textit{the statistical ensemble is a dynamic system},
even if its \textit{elements are stochastic systems}. Use of the statistical
ensemble as some means for calculation of statistical averages is not
necessary. Such an approach may be qualified as the dynamic conception of
statistical description (DCSD). It is appropriate in any case (relativistic
and nonrelativistic). Interpretation of DCSD is carried out in terms of the
ideal fluid and world lines of its particles. Construction of DCSD is a
result of overcoming of the second obstacle.

Between DCSD and axiomatic conception of quantum phenomena (ACQP), i.e.
conventional quantum mechanics, there is a connection. To obtain this
connection it was necessary to obtain fundamental objects of ACQP (wave
function and spin) as attributes of ideal fluid, which appears in DCSD.
Connection between the fluid and the Schr\"{o}dinger equation is known since
the beginning of the quantum mechanics construction \cite{M26,B26}. In after
years many authors developed this interplay known as hydrodynamic
interpretation of quantum mechanics \cite{B52,T52,T53,JZ63,JZ64,HZ69,
B73,BH89,H93}. But this interpretation was founded ultimately on the wave
function as a fundamental object of dynamics. It cannot go outside the
framework of quantum principles, because the connection between the
hydrodynamic interpretation and the quantum mechanics was one-way
connection. One could obtain the irrotational fluid flow from the dynamic
equation for the wave function (Schr\"{o}dinger equation), but one did not
know how to transform dynamic equations for a fluid to the dynamic equation
for a wave function. In other words, we did not know how to describe
rotational fluid flow in terms of the wave function. In terms of the wave
function we could describe only irrotational fluid flow.

To describe arbitrary fluid flow in terms of a wave function, one needs 
\textit{to integrate conventional dynamic equations for a fluid} (Euler
equations). Indeed, the Schr\"{o}dinger equation 
\begin{equation}
i\hbar \frac{\partial \psi }{\partial t}+\frac{\hbar ^{2}}{2m}\mathbf{\nabla 
}^{2}\psi =0  \label{a0.1}
\end{equation}
may be reduced to the hydrodynamic equations for the density $\rho $ and the
velocity $\mathbf{v}$ of some quantum fluid. Substituting $\psi =\sqrt{\rho }%
\exp \left( i\hbar \varphi \right) $ in (\ref{a0.1}) and separating real and
imaginary parts of the equation, we obtain expressions for time derivatives $%
\partial _{0}\rho $ and $\partial _{0}\varphi $. To obtain expression for
the time derivative $\partial _{0}\mathbf{v}$ of the velocity $\mathbf{v=}%
\frac{\hbar }{m}\mathbf{\nabla }\varphi $, we need to differentiate dynamic
equation for $\partial _{0}\varphi $, forming combination $\partial _{0}%
\mathbf{v=\nabla }\left( \frac{\hbar }{m}\partial _{0}\varphi \right) $. The
reverse transition from hydrodynamic equations to dynamic equations for the
wave function needs a general integration of hydrodynamic equations. This
integration is simple in the partial case of irrotational flow, but it is a
rather complicated mathematical problem in the general case, when a result
of integration has to contain three arbitrary functions of three arguments.
Without producing this integration, one cannot derive description of a fluid
in terms of the wave function, and one cannot manipulate dynamic equations,
transforming them from representation in terms of $\rho $, $\mathbf{v}$ to
representation in terms of wave function and back. This problem has not been
solved for years. It had been solved in the end of eighties, and the first
application of this integration can be found in \cite{R89}

Statistical ensemble of discrete dynamic or stochastic systems is a
continuous dynamic system, i.e. some ideal fluid. Integration of
hydrodynamic equations allows one to show that the wave function and spin
are a way of description of an ideal fluid \cite{R99}. This was an
overcoming of the third obstacle. In other words, the wave function appears
as a property of some model (but not as a fundamental object whose
properties are defined by axiomatics). Under some conditions the
irrotational flow of the statistical ensemble (fluid) is described by the
Schr\"{o}dinger equation \cite{R95,R002}. Thus, one can connect MCQP with
the conventional quantum description and show that in the nonrelativistic
case the description in terms of MCQP\ agrees with description in terms of
the conventional quantum mechanics.

Note that the said obstacles were overcame in other order, than they are
listed above. It took about thirty years for overcoming the first obstacle
and construction of T-geometry. In the first paper \cite{R62} the system of
differential equations for the world function of Riemannian geometry \cite
{S60} was obtained. These equations do not contain metric tensor. They
contain only the world function and their derivatives with respect to both
arguments. These equations put the following question. Let the world
function do not satisfy these equations. What is then? Do we obtain
non-Riemannian geometry, or does no geometry exist? Then we could not answer
this question, because as well as other scientists, we believed that the
straight line must be one-dimensional geometrical object (curve) in any
geometry. Only thirty years later we succeeded to answer this question and
to construct non-Riemannian geometry (T-geometry) \cite{R90}. It appeared to
be possible, because we had realized that in some geometries the straight
line can be a non-one-dimensional object (surface). Moreover, it is the
general case of geometry, whereas the Riemannian geometry with
one-dimensional straight lines (geodesic) is a special (degenerate) case.
The string theory suggests this idea of non-one-dimensional straight line.
The second obstacle was overcame the first \cite{R71,R72,R973,R73}. The
third obstacle was overcame only in the end of eighties. Finally the first
obstacle was overcame last \cite{R91}.

It is worth to note that all obstacles had been overcame on the strictly
logical foundation, i.e. only logical constructions based on already known
principles of classical physics. New hypotheses and principles were not
used, and this is not characteristic for the microcosm physics, developed in
XXth century. In other words, the possibility of the MCQP construction was
contained in principles of classical physics. It was necessary only to use
them consistently. Unfortunately, it was unwarranted additional suppositions
and absence of logical consistency, that was the stumbling block for
construction of MCQP.

In the next two sections we consider modification of the quantum phenomena
theory, generated by the overcoming of the said obstacles.

\section{Geometry}

There are two different approaches to geometry: mathematical and physical
ones. In the mathematical approach a geometry is a logical construction
founded on a system of axioms about points and straights. Practically any
system of axioms, containing concepts of a point and of a straight, may be
called a geometry. Well known mathematician Felix Klein \cite{K37} supposed
that only such a construction on a point set is a geometry, where all points
of the set have the same properties (homogeneous geometry). For instance,
Felix Klein insisted that Euclidean geometry and Lobachevsky geometry are
geometries, because they are homogeneous, whereas the Riemannian geometries
are not geometries at all. As a rule the Riemannian geometries are not
homogeneous, and their points have different properties. According to the
Felix Klein viewpoint, they should be called as ''Riemannian topographies''
or as ''Riemannian geographies''. It is a matter of habit and taste how to
call the geometry. But Felix Klein was quite right in the relation, that he
suggested to differ between the Euclidean geometry and the Riemannian one.
The fact is that the principle of the Riemannian geometry construction is
quite different from that of the Euclidean geometry construction. The
Euclidean geometry is constructed on the basis of axioms, whereas the
Riemannian geometry is constructed as a deformation of the Euclidean
geometry.

At the physical approach the geometry is the science on mutual disposition
of points and geometric objects in the space, or events in the space-time.
The mutual disposition is described by the metric $\rho $ (distance between
two points), or by the world function $\sigma =\frac{1}{2}\rho ^{2}$ \cite
{S60}. It is question of the secondary importance, whether all points have
the same properties or not, and what axioms are satisfied by the metric.

The Riemannian geometry is obtained as a result of the proper Euclidean
geometry deformation, when the infinitesimal Euclidean interval $ds_{\mathrm{%
E}}^{2}$ is replaced by the Riemannian interval $ds^{2}=g_{ik}dx^{i}dx^{k}$.
Such a change is a deformation of the Euclidean space. Such an approach to
geometry, when a geometry is a result of the proper Euclidean geometry
deformation will be referred to as a physical approach to geometry. The
obtained geometry will be referred to as a physical geometry. The physical
geometry has no own axiomatics. It uses ''deformed'' Euclidean axiomatics.
The physical geometry describes mutual disposition of points in the space,
or of events in the space-time. It is described by setting the distance
between any two points. The metric $\rho $ is the only characteristic of a
physical geometry. The world function $\sigma =\frac{1}{2}\rho ^{2}$ is more
convenient for description of the physical geometry, because it is real even
for the space-time, where $\rho =\sqrt{2\sigma }$ may be imaginary.

Construction of any physical geometry is determined by the \textit{%
deformation principle} \cite{R02}. It works as follows. The proper Euclidean
geometry $\mathcal{G}_{\mathrm{E}}$ can be described in terms and only in
terms of the world function $\sigma _{\mathrm{E}}$, provided $\sigma _{%
\mathrm{E}}$ satisfies some constraints formulated in terms of $\sigma _{%
\mathrm{E}}$ \cite{R02}. It means that all geometric objects $\mathcal{O}_{%
\mathrm{E}}$ can be described $\sigma $-immanently (i.e. in terms of $\sigma
_{\mathrm{E}}$ and only of $\sigma _{\mathrm{E}}$) $\mathcal{O}_{\mathrm{E}}=%
\mathcal{O}_{\mathrm{E}}\left( \sigma _{\mathrm{E}}\right) $. Relations
between geometric objects are described $\sigma $-immanently by some
expressions $\mathcal{R}_{\mathrm{E}}=\mathcal{R}_{\mathrm{E}}\left( \sigma
_{\mathrm{E}}\right) $. Any physical geometry $\mathcal{G}_{\mathrm{A}}$ can
be obtained from the proper Euclidean geometry by means of a deformation,
when the Euclidean world function $\sigma _{\mathrm{E}}$ is replaced by some
other world function $\sigma _{\mathrm{A}}$ in all definitions of Euclidean
geometric objects $\mathcal{O}_{\mathrm{E}}=\mathcal{O}_{\mathrm{E}}\left(
\sigma _{\mathrm{E}}\right) $ and in all Euclidean relations $\mathcal{R}_{%
\mathrm{E}}=\mathcal{R}_{\mathrm{E}}\left( \sigma _{\mathrm{E}}\right) $
between them. As a result we have the following change 
\[
\mathcal{O}_{\mathrm{E}}=\mathcal{O}_{\mathrm{E}}\left( \sigma _{\mathrm{E}%
}\right) \rightarrow \mathcal{O}_{\mathrm{A}}=\mathcal{O}_{\mathrm{E}}\left(
\sigma _{\mathrm{A}}\right) ,\qquad \mathcal{R}_{\mathrm{E}}=\mathcal{R}_{%
\mathrm{E}}\left( \sigma _{\mathrm{E}}\right) \rightarrow \mathcal{R}_{%
\mathrm{A}}=\mathcal{R}_{\mathrm{E}}\left( \sigma _{\mathrm{A}}\right) 
\]
The set of all geometric objects $\mathcal{O}_{\mathrm{A}}$ and all
relations $\mathcal{R}_{\mathrm{A}}$ between them forms a physical geometry,
described by the world function $\sigma _{\mathrm{A}}$. Index '$\mathrm{E}$'
in the relations of physical geometry $\mathcal{G}_{\mathrm{A}}$ means that
axiomatics of the proper Euclidean geometry was used for construction of
geometric objects $\mathcal{O}_{\mathrm{E}}=\mathcal{O}_{\mathrm{E}}\left(
\sigma _{\mathrm{E}}\right) $ and of relations between them $\mathcal{R}_{%
\mathrm{E}}=\mathcal{R}_{\mathrm{E}}\left( \sigma _{\mathrm{E}}\right) $.
The same axiomatics is used for all geometric objects $\mathcal{O}_{\mathrm{A%
}}=\mathcal{O}_{\mathrm{E}}\left( \sigma _{\mathrm{A}}\right) $ and
relations between them $\mathcal{R}_{\mathrm{A}}=\mathcal{R}_{\mathrm{E}%
}\left( \sigma _{\mathrm{A}}\right) $ in the geometry $\mathcal{G}_{\mathrm{A%
}}$. But now this axiomatics has another form, because of deformation $%
\sigma _{\mathrm{E}}\rightarrow \sigma _{\mathrm{A}}$. It means that the
proper Euclidean geometry $\mathcal{G}_{\mathrm{E}}$ is the basic geometry
for all physical geometries $\mathcal{G}$ obtained by means of a deformation
of the proper Euclidean geometry. If the basic geometry is fixed (it is this
case that will be considered further), the geometry on the arbitrary set $%
\Omega $ of points is called T-geometry (tubular geometry). The T-geometry
is determined \cite{R90,R01} by setting the world function $\sigma $: 
\begin{equation}
\sigma :\;\;\;\Omega \times \Omega \rightarrow \Bbb{R},\qquad \sigma \left(
P,P\right) =0,\qquad \forall P\in \Omega  \label{a1}
\end{equation}
In general, no other constraints are imposed, although one can impose any
additional constraints to obtain a special class of T-geometries. T-geometry
is symmetric, if in addition 
\begin{equation}
\sigma \left( P,Q\right) =\sigma \left( Q,P\right) ,\qquad \forall P,Q\in
\Omega  \label{a1.1}
\end{equation}

Consequent application of \textit{only deformation principle} admits one to
obtain any physical geometry (T-geometry), which appears to be automatically
as consistent as the Euclidean geometry, which lies in its foundation. The
Riemannian geometries form a special class of T-geometries, determined by
the constraint, imposed on the world function 
\begin{equation}
\sigma _{\mathrm{R}}\left( x,x^{\prime }\right) =\frac{1}{2}\left(
\int\limits_{\mathcal{L}_{\left[ xx^{\prime }\right] }}\sqrt{%
g_{ik}dx^{i}dx^{k}}\right) ^{2}  \label{c1}
\end{equation}
where $\sigma _{\mathrm{R}}$ is the world function of the Riemannian space,
and $\mathcal{L}_{\left[ xx^{\prime }\right] }$ means a geodesic segment
between the points $x$ and $x^{\prime }$. Riemannian geometry is determined
by the dimension $n$ and $n\left( n+1\right) /2$ functions $g_{ik}$ of one
point $x$, whereas the class of all possible T-geometries is essentially
more powerful, because it is determined by one function $\sigma $ of two
points $x $ and $x^{\prime }$.

In general, the deformation principle allows one to obtain such geometries,
where non-one-dimensional tubes play the role of the straight lines. The
real space-time geometry is of such a kind. But creators of the Riemannian
geometry supposed that a geometry with tubes instead of straights was
impossible. The restriction (\ref{c1}) on Riemannian geometries was
introduced to forbid deformation transforming one-dimensional Euclidean
straights to many-dimensional tubes. But in nonhomogeneous physical geometry
one fails to suppress the tubular character of straights. In the Riemannian
geometry one succeeds to make this, only refusing from the consequent
application of the deformation principle and using additional means of the
geometry construction. As a result the Riemannian geometry appears to be not
quite consequent construction. This is displayed, in particular, in the lack
of absolute parallelism, whereas in any physical geometry constructed in
accordance with the deformation principle the absolute parallelism takes
place.(see details in \cite{R02}).

The tubular character of timelike straights in the real space-time generates
the stochastic character of the free particles motion in such a space-time,
because the straight (tube) $\mathcal{T}_{P_{0}P_{1}}$, passing through the
points $P_{0}$ and $P_{1}$, is determined by the relation 
\begin{equation}
\mathcal{T}_{P_{0}P_{1}}=\left\{ R|\overrightarrow{P_{0}P_{1}}||%
\overrightarrow{P_{0}R}\right\} =\left\{ R|S_{P_{0}P_{1}R}=0\right\}
\label{s2.1}
\end{equation}
where $\overrightarrow{P_{0}P_{1}}||\overrightarrow{P_{0}R}$ means that
vectors $\overrightarrow{P_{0}P_{1}}$ and $\overrightarrow{P_{0}R}$ are
collinear. In the proper Euclidean space the vectors $\overrightarrow{%
P_{0}P_{1}}$ and $\overrightarrow{P_{0}R}$ are linear dependent (collinear),
if and only if the second order Gram determinant $F_{2}\left(
P_{0},P_{1},R\right) $ vanishes. 
\begin{equation}
\overrightarrow{P_{0}P_{1}}||\overrightarrow{P_{0}R}:\qquad F_{2}\left(
P_{0},P_{1},R\right) =\left| 
\begin{array}{cc}
\left( \overrightarrow{P_{0}P_{1}}.\overrightarrow{P_{0}P_{1}}\right) & 
\left( \overrightarrow{P_{0}P_{1}}.\overrightarrow{P_{0}R}\right) \\ 
\left( \overrightarrow{P_{0}R}.\overrightarrow{P_{0}P_{1}}\right) & \left( 
\overrightarrow{P_{0}R}.\overrightarrow{P_{0}R}\right)
\end{array}
\right| =0  \label{s2.2}
\end{equation}
Here $\left( \overrightarrow{P_{0}P_{1}}.\overrightarrow{P_{0}R}\right) $
denotes the scalar product of two vectors $\overrightarrow{P_{0}P_{1}}$ and $%
\overrightarrow{P_{0}R}$, which is defined by the relation 
\begin{equation}
\left( \overrightarrow{P_{0}P_{1}}.\overrightarrow{P_{0}R}\right) \equiv
\sigma \left( P_{0},P_{1}\right) +\sigma \left( P_{0},R\right) -\sigma
\left( P_{1},R\right)  \label{s2.3}
\end{equation}

Relations (\ref{s2.2}), (\ref{s2.3}) express the collinearity condition via
the world function $\sigma $ of the proper Euclidean space. In other words,
the collinearity of $\overrightarrow{P_{0}P_{1}}$ , $\overrightarrow{P_{0}R}$
is defined $\sigma $-immanently. By definition this relation can be used in
arbitrary T-geometry, i.e. for any world function $\sigma $. The quantity $%
S_{P_{0}P_{1}R}$ is the area of Euclidean triangle with vertices at points $%
P_{0},P_{1},R$. It is connected with the Gram determinant by the relation 
\begin{equation}
F_{2}\left( P_{0},P_{1},R\right) =\left( 2S_{P_{0}P_{1}R}\right) ^{2}
\label{s2.4}
\end{equation}
Thus, two relations (\ref{s2.1}) contain two equivalent conditions of
collinearity.

It is worth to note that conventionally the concept of linear dependence
(collinearity) is introduced in the framework of linear space. To introduce
the linear space, one needs a set of restrictions (fixed dimension,
continuity, coordinate system, etc.). It appears unexpectedly that all these
restrictions are not necessary. The concept of linear dependence can be
introduced on arbitrary set of points, where the world function (metric) is
given. The concept of linear dependence appears to be independent of whether
or not the linear space can be introduced on this set. Concept of linear
dependence describes some very general geometric property, which concerns
the linear space only indirectly.

According to this definition the tube $\mathcal{T}_{P_{0}P_{1}}$ is a set of
such points $R$, that vectors $\overrightarrow{P_{0}P_{1}}$ and $%
\overrightarrow{P_{0}R}$ are collinear. The tubular character of the
straight (thick straight) means that there are many directions $%
\overrightarrow{P_{0}R}$, parallel to the vector $\overrightarrow{P_{0}P_{1}}
$. On the other hand, the motion of a free particle in the curved space-time
is described by the equation of a geodesic 
\begin{equation}
d\dot{x}^{i}=-\Gamma _{kl}^{i}\dot{x}^{k}dx^{l},\qquad dx^{l}=\dot{x}%
^{i}d\tau  \label{a0}
\end{equation}
where $\Gamma _{kl}^{i}$ is the Christoffel symbol. Equations (\ref{a0})
describe the parallel transport of the velocity vector $\dot{x}^{i}$ of the
particle along the direction $dx^{i}=\dot{x}^{i}d\tau $, determined by the
velocity vector $\dot{x}^{i}$. If there are many vectors parallel to the
velocity vector $\dot{x}^{i}$, the parallel transport (\ref{a0}) appears to
be not single-valued, and the world line becomes to be random.

The flat uniform isotropic space-time is described by the world function 
\cite{R91} 
\begin{equation}
\sigma =\sigma _{\mathrm{M}}+D\left( \sigma _{\mathrm{M}}\right) ,\qquad
D\left( \sigma _{\mathrm{M}}\right) =\frac{\hbar }{2bc}\geq 10^{-21}\text{cm}%
^{2},\quad \text{if}\quad \sigma _{\mathrm{M}}>\frac{\hbar }{2bc}
\label{a2.1}
\end{equation}
where $\sigma _{\mathrm{M}}$ is the world function of the Minkowski space, $%
c $ is the speed of the light. The distortion function $D\left( \sigma _{%
\mathrm{M}}\right) $ describes the character of quantum stochasticity. In
the space with nonvanishing distortion $D\left( \sigma _{\mathrm{M}}\right) $
the particle mass is geometrized \cite{R91}. The quantity $b\leq 10^{-17}$%
g/cm is the constant, describing connection between the geometric mass $\mu $
and the usual mass $m$ by means of the relation $m=b\mu $. Form of the
distortion function $D\left( \sigma _{\mathrm{M}}\right) $ is determined by
the demand that the stochasticity generated by distortion is the quantum
stochasticity, i.e. the statistical description of the free stochastic
particle motion is equivalent to the quantum description in terms of the
Schr\"{o}dinger equation \cite{R91}.

\section{Statistical description}

Let the statistical ensemble $\mathcal{E}_{\mathrm{d}}\left[ \mathcal{S}_{%
\mathrm{d}}\right] $ of deterministic classical particles $\mathcal{S}_{%
\mathrm{d}}$ be described by the action $\mathcal{A}_{\mathcal{E}_{\mathrm{d}%
}\left[ \mathcal{S}_{\mathrm{d}}\left( P\right) \right] }$, where $P$ are
parameters describing $\mathcal{S}_{\mathrm{d}}$ (for instance, mass,
charge). Let under influence of some stochastic agent the deterministic
particle $\mathcal{S}_{\mathrm{d}}$ turn to a stochastic particle $\mathcal{S%
}_{\mathrm{st}}$. The action $\mathcal{A}_{\mathcal{E}_{\mathrm{st}}\left[ 
\mathcal{S}_{\mathrm{st}}\right] }$ for the statistical ensemble $\mathcal{E}%
_{\mathrm{st}}\left[ \mathcal{S}_{\mathrm{st}}\right] $ of stochastic
particles $\mathcal{S}_{\mathrm{st}}$ is reduced to the action $\mathcal{A}_{%
\mathcal{S}_{\mathrm{red}}\left[ \mathcal{S}_{\mathrm{d}}\right] }=\mathcal{A%
}_{\mathcal{E}_{\mathrm{st}}\left[ \mathcal{S}_{\mathrm{st}}\right] }$ for
some set $\mathcal{S}_{\mathrm{red}}\left[ \mathcal{S}_{\mathrm{d}}\right] $
of identical interacting deterministic particles $\mathcal{S}_{\mathrm{d}}$.
The action $\mathcal{A}_{\mathcal{S}_{\mathrm{red}}\left[ \mathcal{S}_{%
\mathrm{d}}\right] }$ as a functional of $\mathcal{S}_{\mathrm{d}}$ has the
form $\mathcal{A}_{\mathcal{E}_{\mathrm{d}}\left[ \mathcal{S}_{\mathrm{d}%
}\left( P_{\mathrm{eff}}\right) \right] }$, where parameters $P_{\mathrm{eff}%
}$ are parameters $P$ of the deterministic particle $\mathcal{S}_{\mathrm{d}%
} $, averaged over the statistical ensemble, and this averaging describes
interaction of particles $\mathcal{S}_{\mathrm{d}}$ in the set $\mathcal{S}_{%
\mathrm{red}}\left[ \mathcal{S}_{\mathrm{d}}\right] $ \cite{R002c,R003}. It
means that 
\begin{equation}
\mathcal{A}_{\mathcal{E}_{\mathrm{st}}\left[ \mathcal{S}_{\mathrm{st}}\right]
}=\mathcal{A}_{\mathcal{S}_{\mathrm{red}}\left[ \mathcal{S}_{\mathrm{d}%
}\left( P\right) \right] }=\mathcal{A}_{\mathcal{E}_{\mathrm{d}}\left[ 
\mathcal{S}_{\mathrm{d}}\left( P_{\mathrm{eff}}\right) \right] }
\label{a0.6a}
\end{equation}
In other words, stochasticity of particles $\mathcal{S}_{\mathrm{st}}$ in
the ensemble $\mathcal{E}_{\mathrm{st}}\left[ \mathcal{S}_{\mathrm{st}}%
\right] $ is replaced by interaction of $\mathcal{S}_{\mathrm{d}}$ in $%
\mathcal{S}_{\mathrm{red}}\left[ \mathcal{S}_{\mathrm{d}}\right] $, and this
interaction is described by a change 
\begin{equation}
P\rightarrow P_{\mathrm{eff}}  \label{a0.6b}
\end{equation}
in the action $\mathcal{A}_{\mathcal{E}_{\mathrm{d}}\left[ \mathcal{S}_{%
\mathrm{d}}\left( P\right) \right] }$.

Action for the statistical ensemble of free deterministic particles has the
form 
\begin{equation}
\mathcal{A}\left[ x\right] =-\int mc\sqrt{g_{ik}\dot{x}^{i}\dot{x}^{k}}d\xi
_{0}d\vec{\xi},\qquad \dot{x}^{i}\equiv \frac{dx^{i}}{d\xi _{0}}
\label{b0.16}
\end{equation}
where $x=\left\{ x^{i}\right\} $, $i=0,1,2,3$ is a function of $\xi =\left\{
\xi _{0},\vec{\xi}\right\} =\left\{ \xi _{i}\right\} ,$ $\;i=0,1,2,3$.

The only parameter for the free particle is its mass $m$, and the change (%
\ref{a0.6b}) in the nonrelativistic case has the form 
\begin{equation}
m\rightarrow m_{\mathrm{eff}}=m\left( 1-\frac{\mathbf{u}^{2}}{2c^{2}}+\frac{%
\hbar }{2mc^{2}}\mathbf{\nabla u}\right)  \label{a0.9a}
\end{equation}
where $\mathbf{u}=\mathbf{u}\left( t,\mathbf{x}\right) $ is the mean value
of the stochastic velocity component. Quantum constant $\hbar $ appears here
as coupling constant between the regular and stochastic components of the
particle velocity. The velocity $\mathbf{u}$ is considered to be a new
dependent variable, and dynamic equation for $\mathbf{u}$ is obtained as a
result of the action variation with respect to $\mathbf{u}$ \cite{R003}. The
velocity $\mathbf{u}$ is supposed to be small as compared with the speed of
the light $c$.

In the relativistic case the change (\ref{a0.6b}) takes the form 
\begin{equation}
m^{2}\rightarrow m_{\mathrm{eff}}^{2}=m^{2}\left( 1+u_{l}u^{l}+\lambda
\partial _{l}u^{l}\right) ,\qquad \lambda =\frac{\hbar }{mc}  \label{c2.1}
\end{equation}
where $u^{l}=\left\{ u^{0},\mathbf{u}\right\} $. Then the action (\ref{b0.16}%
) is transformed to the form 
\begin{equation}
\mathcal{A}\left[ x,\kappa \right] =-\int mcK\sqrt{g_{ik}\dot{x}^{i}\dot{x}%
^{k}}d\xi _{0}d\vec{\xi} ,\qquad K=\sqrt{1+\frac{\hbar ^{2}}{m^{2}c^{2}}%
\left( \kappa ^{l}\kappa _{l}+\partial _{l}\kappa ^{l}\right) }
\label{a0.16}
\end{equation}
where dependent variables $x=\left\{ x^{i}\right\} $, $i=0,1,2,3$ are a
function of variables $\xi =\left\{ \xi _{0},\vec{\xi}\right\} =\left\{ \xi
_{i}\right\} ,$ $\;i=0,1,2,3$. Dependent variables $\kappa =\left\{ \kappa
^{i}\right\} ,$ $\;i=0,1,2,3$ are functions of $x$. The metric tensor $%
g_{ik}=$diag$\left\{ c^{2},-1,-1,-1\right\} $. Variables $\kappa ^{l}$ are
connected with $u^{l}$ by means of the relation 
\begin{equation}
u^{l}=\frac{\hbar }{m}\kappa ^{l},\qquad l=0,1,2,3  \label{a0.17}
\end{equation}

On one hand, the action (\ref{a0.16}) describes a set of deterministic
particles interacting between themselves via self-consistent vector field $%
\kappa ^{l}$. On the other hand, the action (\ref{a0.16}) describes a
quantum fluid. Irrotational flow of this fluid is described by one-component
wave function $\psi $, satisfying the Klein-Gordon equation \cite{R98,R003}.

In general case the fluid flow is described by two-component wave function,
satisfying the dynamic equation \cite{R003} 
\begin{equation}
-\hbar ^{2}\partial _{k}\partial ^{k}\psi -\left( m^{2}c^{2}+\frac{\hbar ^{2}%
}{4}\left( \partial _{l}s_{\alpha }\right) \left( \partial ^{l}s_{\alpha
}\right) \right) \psi =\hbar ^{2}\frac{\partial _{l}\left( \rho \partial
^{l}s_{\alpha }\right) }{2\rho }\left( \sigma _{\alpha }-s_{\alpha }\right)
\psi  \label{A.46}
\end{equation}
where 3-vector $\mathbf{s=}\left\{ s_{1},s_{2},s_{3},\right\} $ is
determined by the relations 
\begin{equation}
\psi =\left( _{\psi _{2}}^{\psi _{1}}\right) ,\qquad \psi ^{\ast }=\left(
\psi _{1}^{\ast },\psi _{2}^{\ast }\right) ,\qquad \rho =\psi ^{\ast }\psi
,\qquad s_{\alpha }=\frac{\psi ^{\ast }\sigma _{\alpha }\psi }{\rho },\qquad
\alpha =1,2,3  \label{A.42}
\end{equation}
Here $\vec{\sigma}=\left\{ \sigma _{1},\sigma _{2},\sigma _{3}\right\} $ are
Pauli matrices.

From physical viewpoint the quantization procedure (\ref{c2.1}), when the
mass $m$ is replaced by its mean value $m_{\mathrm{eff}}$, looks rather
reasonable. Indeed, the value of $m_{\mathrm{eff}}$ depends on the state of
the stochastic velocity component. Stochastic velocity component has
infinite number of the freedom degrees and consideration of influence of the
mean value $u^{l}$ on the regular component of the particle velocity appears
to be very complicated. It is described by partial differential equations,
whereas in absence of this influence the regular particle motion is
described by the ordinary differential equations. From physical viewpoint
such an interpretation of the quantization looks more reasonable, than
conventional interpretation in terms of wave function and operators.

The wave function and spin appear here as a way of description of the ideal
fluid \cite{R99}, i.e. as fluid attributes. In other words, statistical
description and hydrodynamic interpretation of the world function are
primary, and wave function is secondary. Hierarchy of concepts is described
by two schemes in figure 2. Note that transition from ACQP to MCQP became
possible only after solution of such a pure mathematical problem as
integration of complete system of dynamic equations for ideal fluid.
Systematical application of this integration for description of quantum
phenomena began in 1995 \cite{R95,R995}.

\section{Methods and capacities of MCQP}

Let us formulate the first difference between ACQP and MCQP. As any
statistical theory MCQP contains two sorts of investigated objects:
stochastic system $\mathcal{S}_{\mathrm{st}}$ and statistically averaged
system $\left\langle \mathcal{S}_{\mathrm{st}}\right\rangle $, whereas ACQP
investigates only one sort of objects: so called quantum systems $\mathcal{S}%
_{\mathrm{q}}$. Systems $\left\langle \mathcal{S}_{\mathrm{st}}\right\rangle 
$ and $\mathcal{S}_{\mathrm{q}}$ are continuous dynamic systems. They have
dynamic equations and coincide practically always. The system $\mathcal{S}_{%
\mathrm{st}}$ is a discrete stochastic system, for which dynamic equations
are absent. Dynamic system $\left\langle \mathcal{S}_{\mathrm{st}%
}\right\rangle $ appears as a result of statistical description of $\mathcal{%
S}_{\mathrm{st}}$ (consideration of statistical ensemble $\mathcal{E}\left[ 
\mathcal{S}_{\mathrm{st}}\right] $). The statistically averaged system $%
\left\langle \mathcal{S}_{\mathrm{st}}\right\rangle $ is the statistical
ensemble $\mathcal{E}\left[ \mathcal{S}_{\mathrm{st}}\right] $, normalized
to one system, and $\left\langle \mathcal{S}_{\mathrm{st}}\right\rangle $ as
a continuous dynamic system have practically all properties of statistical
ensemble. Normalization to one system is possible, because properties of the
statistical ensemble $\mathcal{E}\left[ N,\mathcal{S}_{\mathrm{st}}\right] $
do not depend on number $N$ of elements in it, if $N$ is large enough. Then
one can set formally $N=1$, and $\mathcal{E}\left[ N,\mathcal{S}_{\mathrm{st}%
}\right] $ turns to $\left\langle \mathcal{S}_{\mathrm{st}}\right\rangle $.
But $\left\langle \mathcal{S}_{\mathrm{st}}\right\rangle $ remains to be
continuous dynamic system and conserves all properties of the statistical
ensemble $\mathcal{E}\left[ \infty ,\mathcal{S}_{\mathrm{st}}\right] $ \cite
{R002c}.

Two kinds of measurement in MCQP ($S$-measurement and $M$-measurement) is
another side of existence of two sorts of objects $\mathcal{S}_{\mathrm{st}}$
and $\left\langle \mathcal{S}_{\mathrm{st}}\right\rangle $. Single
measurement ($S$-measurement) is produced over single stochastic system $%
\mathcal{S}_{\mathrm{st}}$. Result of $S$-measurement is random. It means
that the result of $S$-measurement is not reproduced, in general, in other $%
S $-measurements on other $\mathcal{S}_{\mathrm{st}}$, prepared in the same
way. $S$-measurement always leads to one random but definite value $%
R^{\prime }$ of the measured quantity $\mathcal{R}$. This result does not
influence on the wave function $\psi $, because wave function describes the
state of $\left\langle \mathcal{S}_{\mathrm{st}}\right\rangle $ and has no
relation to $\mathcal{S}_{\mathrm{st}}$.

The massive measurement ($M$-measurement) is produced over many single
dynamic systems $\mathcal{S}_{\mathrm{st}}$, constituting $\left\langle 
\mathcal{S}_{\mathrm{st}}\right\rangle $. $M$-measurement is a set of many $%
S $-measure\-ments, produced over different elements of $\mathcal{E}\left[ 
\mathcal{S}_{\mathrm{st}}\right] $ or of $\left\langle \mathcal{S}_{\mathrm{%
st}}\right\rangle $. $M$-measurement of the quantity $\mathcal{R}$ leads
always to a distribution $F\left( R^{\prime }\right) $ of possible values $%
R^{\prime }$, because $S$-measurements in different elements of $%
\left\langle \mathcal{S}_{\mathrm{st}}\right\rangle $ give different
results, in general. Result of $M$-measurement is always a distribution,
even in the case, when all $S$-measurements, constituting the $M$%
-measurement, give the same result. In the last case the distribution is a $%
\delta $-function. Distribution $F\left( R^{\prime }\right) $ is
reproducible. It is reproduced in other $M$-measurements, carried out in the
same state $\psi $.

Thus, $S$-measurement is irreproducible, in general, and gives a definite
result. The $M$-measurement is reproducible and, in general, does not give a
definite result (but only a distribution of results). Nevertheless, formally
the $M$-measurement of the quantity $\mathcal{R}$ may be considered to be a
single act of measurement of the quantity $R^{\prime }$ distribution, which
is produced over the dynamic system $\left\langle \mathcal{S}_{\mathrm{st}%
}\right\rangle $.

Theory can predict only results of $M$-measurement, which leads to a
reproducible distribution except for the case, when the measured system is
an element of the statistical ensemble, which is found in such a state,
where the distribution of the measured quantity is a $\delta $-function. In
this case the result of $S$-measurement can be predicted, because it is
determined uniquely by the distribution, which is a result of $M$%
-measurement and can be predicted by the theory.

Is it possible that $M$-measurement of the quantity $\mathcal{R}$ in the
state $\psi $ lead to a definite result $R^{\prime }$? Yes, it is possible,
if $M$-measurement is accompanied by a selection of those $S$-measurements,
constituting the $\ M$-measurement, which give the result $R^{\prime }$.
Such a measurement is called selective $M$-measurement (or $SM$%
-measurement). $SM$-measurement has properties of $S$-measurement (definite
result $R^{\prime }$) and those of $M$-measurement (it is produced in $%
\left\langle \mathcal{S}_{\mathrm{st}}\right\rangle $, and influences on the
state of $\left\langle \mathcal{S}_{\mathrm{st}}\right\rangle $). $SM$%
-measurement is accompanied by a selection, and the question about its
reproducibility is meaningless, because it depends on the kind of selection,
which determines the result of the $SM$-measurement.

In ACQP there is only one type of objects (quantum systems $\mathcal{S}_{%
\mathrm{q}}$) and only one type of measurement ($Q$-measurement). The $Q$%
-measurement is essentially $M$-measurement, because it always gives
distribution of the measured quantity $\mathcal{R}$. Result of $Q$%
-measurement is determined by the rule of von Neumann. All predictions in
ACQP as well as in MCQP concern only with results of $M$-measurement. In
ACQP there is only one object and only one type of measurement. Nobody
distinguishes in ACQP between $S$-measurement and $M$-measurement.

In ACQP the statistically averaged system $\left\langle \mathcal{S}_{\mathrm{%
st}}\right\rangle $ is considered to be an individual quantum system $%
\mathcal{S}_{\mathrm{q}}$ and a single measurement on $\mathcal{S}_{\mathrm{q%
}}=\left\langle \mathcal{S}_{\mathrm{st}}\right\rangle $ is considered to be
a measurement with properties of both $M$-measurement and $S$-measurement
(i.e. $SM$-measurement). In classical physics a single physical system is
described as a discrete dynamic system, whereas a statistical ensemble of
single physical systems is described as a continuous dynamic system. In ACQP
the question, why the continuous dynamic system $\mathcal{S}_{\mathrm{q}}$
describes a single physical object (for instance, a single particle), is not
raised usually. If nevertheless such a question is raised, one answers that
quantum system is a special kind of physical system which distinguishes from
classical system and whose state is described by such a mysterious quantity
as wave function, or something like this. The fact that the wave function is
a method of the statistical ensemble description is unknown practically.

In ACQP there is only one object, and for description of a single
measurement one uses $SM$-measurement (but not $S$-measurement), because $S$%
-measurement does not exist in framework of ACQP. The $SM$-measurement is
produced under the dynamic system $\mathcal{S}_{\mathrm{q}}=\left\langle 
\mathcal{S}_{\mathrm{st}}\right\rangle $ and influences on the wave
function, describing the state of $\mathcal{S}_{\mathrm{q}}$. The $S$%
-measurement is produced under the system $\mathcal{S}_{\mathrm{st}}$. It
does not influence on the wave function, describing the state of $\mathcal{S}%
_{\mathrm{q}}=\left\langle \mathcal{S}_{\mathrm{st}}\right\rangle $.
Character of influence on the wave function is the main formal difference
between $S$-measurement and $SM$-measurement.

Replacement of $S$-measurement by $SM$-measurement in ACQP leads to such
paradoxes, as the Schr\"{o}dinger cat paradox, or the EPR-paradox, which are
based on the belief that a single measurement changes the wave function
(i.e. that a single measurement is a $SM$-measurement). Paradoxes are
eliminated, if one take into account that a single measurement is a $S$%
-measurement, which does not change the wave function). See detail
discussion in \cite{R002c}.

Some properties of $\left\langle \mathcal{S}_{\mathrm{st}}\right\rangle $
appear as a result of properties of $\mathcal{S}_{\mathrm{st}}$. Another
ones appear as a result of statistical averaging. The last are interpreted
as collective properties. For instance, in statistical physics (MCTP) the
temperature is a collective property, because one molecule has no
temperature. Only collective of molecules has a temperature in statistical
physics. In thermodynamics (ACTP) there is no concept of collective
properties, and the temperature is a property of any amount of matter (one
molecule, or even a half of molecule). In MCQP the wave function $\psi $ is
a collective quantity, which describes the state of $\left\langle \mathcal{S}%
_{\mathrm{st}}\right\rangle $. If we say that individual stochastic system $%
\mathcal{S}_{\mathrm{st}}$ is found in the state $\psi $, it means only that 
$\mathcal{S}_{\mathrm{st}}$ is taken from the ensemble $\mathcal{E}\left[ 
\mathcal{S}_{\mathrm{st}}\right] $, whose state is described by the wave
function $\psi $. In MCQP the spin of a particle may be a property of
individual particle $\mathcal{S}_{\mathrm{st}}$, and it may be a collective
property of $\left\langle \mathcal{S}_{\mathrm{st}}\right\rangle $. In
different cases we have different results.

In the case of dynamic system $\mathcal{S}_{\mathrm{P}}$ described by the
Pauli equation the electron spin is the collective property \cite{R95}. It
means, in particular, that the dynamic system $\mathcal{S}_{\mathrm{S}}$,
described by the Schr\"{o}dinger equation, and dynamic system $\mathcal{S}_{%
\mathrm{P}}$ consist of similar individual systems $\mathcal{S}_{\mathrm{st}%
} $. The difference between them is described by the type of the fluid flow.
In the case of $\mathcal{S}_{\mathrm{S}}$ the fluid flow is irrotational,
whereas in the case of $\mathcal{S}_{\mathrm{P}}$ the fluid flow is
rotational. In other words, in $\mathcal{S}_{\mathrm{P}}$ the electron spin
is conditioned by collective property (vortical flow). In the given case the
electron spin is a result of the fluid flow vorticity. (Let us remember that
the statistical ensemble is a fluidlike dynamic system.)

In the case of dynamic system $\mathcal{S}_{\mathrm{D}}$, described by the
free Dirac equation, the electron spin is a property of an individual
stochastic particle \cite{R001}. In this case the dynamic disquatization of
the Dirac equation, i.e. determination of classic analog $\mathcal{S}_{%
\mathrm{Dcl}}$ of the Dirac particle $\mathcal{S}_{\mathrm{D}}$ shows that $%
\mathcal{S}_{\mathrm{Dcl}}$ is a dynamic system, having ten degrees of
freedom. It may be interpreted as two classical particles, rotating around
their common center of inertia. Angular momentum of this rotation forms spin
of $\mathcal{S}_{\mathrm{Dcl}}$. Thus, in the case of the Dirac electron $%
\mathcal{S}_{\mathrm{D}}$ the spin is a property of an individual stochastic
system.

In ACQP there are no collective properties, because there is only one
object: the quantum system $\mathcal{S}_{\mathrm{q}}$. The statement of the
question whether the spin of the particle is a collective property is
meaningless in the framework of ACQP. Is it important for investigations to
distinguish between individual and collective properties? Sometimes it is of
no importance, but sometimes it is important. We can investigate this
question in the example of temperature in the statistical physics.
Collective properties of temperature are of no importance, provided we deal
with the mean values of it. If we deal with fluctuations of temperature and
those of other quantities, the collective properties of temperature become
to be important. In any case, method of investigation of MCQP appears to be
more subtle and effective, than that of ACQP.

In general, MCQP as any model conception possesses \textit{more subtle and
flexile methods of investigation}. For instance, the change (\ref{c2.1})
carries out essentially the quantization procedure, i.e. transition from
classical description to the quantum one. Let us imagine that the
quantization (\ref{c2.1}) is not exactly true, and one needs to correct it
slightly. In the framework of MCQP one needs only to change the form of the
expression (\ref{c2.1}). It is not clear how such a modification can be
realized in the framework of ACQP, because ACQP is connected closely with
linearity of dynamic equations in terms of the wave function. One cannot
imagine quantum mechanics which is nonlinear in terms of wave function,
because in this case the quantum principles are violated.

Having a long history \cite{M26,B26}, the hydrodynamic interpretation should
rank among new methods of investigation. But in the framework of ACQP the
hydrodynamic interpretation is secondary as it follows from the schemes in
figure 2. In the framework of MCQP the hydrodynamic interpretation is
primary, and possesses more subtle methods of investigation. In the
framework of ACQP a transition to semiclassical approximation is carried out
by means of transition to the limit $\hbar \rightarrow 0$ with some
additional conditions. In MCQP this transition is realized by means of
dynamic disquantization \cite{R001}, which is a relativistic dynamic
procedure. At the dynamic disquantization one removes transversal components 
$\partial _{\perp k}=\partial _{k}-j_{k}j^{l}\left( j_{s}j^{s}\right)
^{-1}\partial _{l}$ of derivative $\partial _{k}$, which are orthogonal to
the flux 4-vector $j^{k}$. After this transformation the dynamic equations
contain derivatives only in direction of the vector $j^{k}$. Such a system
of partial differential equations can be reduced to a system of ordinary
dynamic equations. As a result of dynamic disquantization the system of
partial differential equations turns to a system of ordinary differential
equations. In other words, the statistical ensemble of stochastic systems
turns to a statistical ensemble of dynamic systems. As a result the
continuous system can be interpreted in terms of a discrete dynamic system
(with finite number of the freedom degrees). In the nonrelativistic case the
dynamic disquantization is equivalent to $\hbar \rightarrow 0$. In the
relativistic case the quantum constant $\hbar $ remains in the discrete
dynamic system \cite{R001}. It allows one to obtain a more subtle
interpretation.

This method was applied for investigation of dynamic system $\mathcal{S}_{%
\mathrm{D}}$, described by the Dirac equation \cite{R001}. It appears that
the classical analog of the Dirac particle $\mathcal{S}_{\mathrm{D}}$ is a
rotator (but not a single particle), i.e. two particles rotating around
their common center of inertia. This explains freely angular and magnetic
momenta of the Dirac particle. Dynamic variables describing rotation contain
the quantum constant $\hbar $. If $\hbar \rightarrow 0$, degrees of freedom,
connected with rotation are suppressed. Radius of rotator tends to $0$, and
instead of rotator we obtain pointlike particle with spin and magnetic
moment. This result with $\hbar \rightarrow 0$ agrees with the results,
obtained in the framework of ACQP. More soft result of rotator, when $\hbar
\neq 0$, cannot be obtained by methods of ACQP. These methods are too rough.
Besides, it appears (quite unexpectedly) that the internal (rotational)
degrees of freedom of the dynamic system $\mathcal{S}_{\mathrm{D}}$ are
described in nonrelativistic manner \cite{R001,R995,R001b}. Investigation of
the dynamic system $\mathcal{S}_{\mathrm{D}}$ was produced without any
additional supposition. It was investigated simply as a dynamic system by
means of relativistically covariant methods. These results cannot be
obtained in the framework of conventional quantum mechanics (ACQP).

Methods of MCQP are consistent relativistic ones. For description of
relativistic quantum phenomena ACQP uses the program of uniting of
nonrelativistic quantum mechanics technique with the relativity principles.
Unfortunately, this program failed. At any rate, it works unsuccessfully
last fifty years, trying to construct relativistic quantum field theory and
theory of elementary particles. It seems that uniting of nonrelativistic
quantum mechanics technique with the relativity principles is impossible.

As another example of the MCQP\ methods application, we refer to the problem
of the pair production, which is the central problem in the high energy
physics. ACQP cannot say anything on the pair production mechanism and on
the agents, responsible for this process, whereas MCQP can say something
pithy on this problem. MCQP vests responsibility for the pair production on
the $\kappa $-field (\ref{a0.17}), which is conditioned by the stochastic
component of the particle motion. In MCQP the pair production is taken into
account on the descriptive (before-dynamic) level, i.e. the pair production
is taken into account by consideration of WL as a primary physical object,
whereas in ACQP the pair production is taken into account only on the
dynamic level, i.e. by means of dynamic equations. In ACQP particles may be
produced not only by pairs. The number of particles produced in an
elementary act may be arbitrary. The number of produced particles depends on
the form of corresponding term in Lagrangian. In MCQP the particles are
produced only by pairs particle - antiparticle. This fact is fixed on the
descriptive (conceptual) level, and cannot be changed on dynamical level (by
a choice of Lagrangian). All this shows that the source of pair production
is different in MCQP and ACQP.

Let us imagine that the particle world line turns in the time direction.
Depending on situation, such a turn describes either pair production, or
pair annihilation. The $\kappa $-field creates conditions for such a turn
and for the pair production. The fact is that at such a turn in time the
world line direction becomes spacelike ($p^{2}=m^{2}<0$) in the vicinity of
the turning point. If one forbids the world line to be spacelike, the pair
production becomes to be impossible. Such a possibility to change the
particle mass and to make it imaginary is rather rare property among the
force fields. For instance, the electromagnetic field of any magnitude
cannot change the particle mass, and hence, to produce pairs. Pair
production is a prerogative of the $\kappa $-field. According to relation (%
\ref{c2.1}) the expression containing $\kappa $-field enter in the effective
squared mass as a factor. If this expression is negative, the mass becomes
imaginary, and the pair production (annihilation) becomes to be possible 
\cite{R003}.

Furthermore, pair production, obtained in ACQP at canonical quantization of
nonlinear relativistic field, does not take place in reality. It was shown
at canonical quantization of nonlinear complex scalar field \cite{R01a},
described by the Lagrangian density 
\begin{equation}
L=:\varphi _{i}^{\ast }\varphi ^{i}-m^{2}\varphi ^{\ast }\varphi +{\frac{%
\lambda }{2}}\varphi ^{\ast }\varphi ^{\ast }\varphi \varphi :  \label{a4.1}
\end{equation}
\[
\varphi =\varphi (x),\qquad \varphi _{i}\equiv \partial _{i}\varphi ,\qquad
\varphi ^{i}\equiv \partial ^{i}\varphi ,\qquad x=(t,x). 
\]
At canonical quantization of (\ref{a4.1}) the $WL$-scheme of quantization
was used, when the object of quantization is WL, i.e. world line considered
as the primary physical object. In the $WL$-scheme the canonical
quantization is produced without imposition of additional constraint

\begin{equation}
\left[ u,P_{0}\right] _{-}=-i\hbar \frac{\partial u}{\partial x^{0}},\qquad
E=P^{0}=\int T^{00}d\mathbf{x}  \label{a1.19}
\end{equation}
where $\left[ ...\right] _{-}$ denotes commutator and $T^{ik}$ is the
energy-momentum tensor. The condition (\ref{a1.19}) identifies the energy
with the evolution operator (Hamiltonian). This identification is possible
in the case, when there are only particles, or only antiparticles. It is
possible also in the case, when particles and antiparticles are considered
as different physical objects (but not as attributes of WL). In $WL$-scheme
of quantization the number of primary physical objects (WLs) is conserved,
and quantization is produced exactly (without the perturbation theory
methods). In such a quantization the pair production is absent, that agrees
with demands to the field producing pairs.

But then the question arises. Why does pair production appear at
quantization according to $PA$-scheme \cite{GJ68,GJ70,GJ970,GJ72}, when
particle and antiparticle are considered as primary objects? The answer is
as follows. Canonical quantization (according to $WL$-scheme) is possible
without imposition of constraint (\ref{a1.19}). It means that the constraint
(\ref{a1.19}) is an additional condition, and one should to verify its
compatibility with dynamic equations. Unfortunately, nobody had verified
this, supposing that (\ref{a1.19}) is a necessary condition of the second
quantization and there is no necessity to verify its compatibility with
dynamic equations. This test was produced in \cite{R01a}. It appears that (%
\ref{a1.19}) is compatible with dynamic equations, provided $\lambda =0$,
i.e. the field is linear. In the nonlinear case $\lambda \neq 0$ imposition
of the constraint (\ref{a1.19}) leads to overdetermination of the problem.
In the overdetermined (and hence, inconsistent) problem one can obtain
practically any results, which one wishes. So, authors of \cite
{GJ68,GJ70,GJ970,GJ72} wanted to obtain the pair production, and they had
obtained it.

These examples show that MCQP and its subtle investigation methods can be
useful at investigation of the microcosm phenomena properties.

Thus, MCQP makes the first successes, but not in the sense that it explains
some new experiments, which could not be explained before. MCQP uses the
more subtle dynamic methods of investigation (consideration of two objects $%
\mathcal{S}_{\mathrm{st}}$ and $\left\langle \mathcal{S}_{\mathrm{st}%
}\right\rangle $, hydrodynamic interpretation of relativistic processes \cite
{R98,R003}, dynamic quantization and disquantization \cite{R001}), which
cannot be used by ACQP because of its axiomatic character. Difference
between the methods of MCQP and ACQP is described by the following scheme

\noindent $
\begin{array}{cc}
\text{ACQP} & \text{MCQP} \\ 
\begin{array}{c}
\text{Combination of nonrelativistic } \\ 
\text{quantum technique with } \\ 
\text{principles of relativity}
\end{array}
& 
\begin{array}{c}
\text{Consequent relativistic description } \\ 
\text{at all stages}
\end{array}
\\ 
\begin{array}{c}
\text{1. Additional hypotheses are used} \\ 
\text{(QM principles) }
\end{array}
& 
\begin{array}{c}
\text{1. \textit{No additional hypotheses are used}}
\end{array}
\\ 
\begin{array}{c}
\text{2. One kind of measurement, as } \\ 
\text{far as only one statistical average } \\ 
\text{object }\left\langle \mathcal{S}\right\rangle \text{ is considered. It
is } \\ 
\text{referred to as a quantum system}
\end{array}
& 
\begin{array}{c}
\text{2. Two kinds of measurement, because } \\ 
\text{two kinds of objects (individual }\mathcal{S}_{\mathrm{st}}\text{ } \\ 
\text{and statistical average }\left\langle \mathcal{S}\right\rangle \text{)
are } \\ 
\text{considered}
\end{array}
\\ 
\begin{array}{c}
\text{3. Quantization: procedure on } \\ 
\text{the conceptual level:} \\ 
\mathbf{p}\rightarrow -i\hbar \mathbf{\nabla }\;\;\;\text{etc. }
\end{array}
& 
\begin{array}{c}
\text{3. Dynamic quantization: relativistic } \\ 
\text{procedure on the dynamic level} \\ 
m^{2}\rightarrow m_{\mathrm{eff}}^{2}=m^{2}+\frac{\hbar ^{2}}{c^{2}}\left(
\kappa _{l}\kappa ^{l}+\partial _{l}\kappa ^{l}\right)
\end{array}
\\ 
\begin{array}{c}
\text{4. Transition to classical description:} \\ 
\text{procedure on conceptual level} \\ 
\hbar \rightarrow 0\qquad \psi \rightarrow \left( \mathbf{x},\mathbf{p}%
\right)
\end{array}
& 
\begin{array}{c}
\text{4. Dynamic disquantization: relativistic} \\ 
\text{ procedure on dynamic level} \\ 
\partial ^{k}\rightarrow \frac{j^{k}j^{l}}{j_{s}j^{s}}\partial _{l}
\end{array}
\\ 
\begin{array}{c}
\text{5. Interpretation in terms of wave} \\ 
\text{function }\psi
\end{array}
& 
\begin{array}{c}
\text{5. Interpretation in terms of statistical} \\ 
\text{average world lines (WL)} \\ 
\frac{dx^{i}}{d\tau }=j^{i}\left( x\right) ,\;\;\; \\ 
j^{k}=-\frac{i\hbar }{2}\left( \psi ^{\ast }\partial ^{k}\psi -\partial
^{k}\psi ^{\ast }\cdot \psi \right)
\end{array}
\end{array}
$

\bigskip

MCQP is essentially more flexible conception, than ACQP, as far as all in
MCQP is determined by the space-time geometry, and the set of all possible
geometries is described by a function of two arguments. This is a great
reserve for corrections and modifications of MCQP. At the same time all
modifications of MCQP are restricted by a change of the world function, and
possible modifications do not concern the structure of MCQP, which is
founded on several principles, connected logically between themselves.

On the contrary ACQP is founded on a set of rigid rules, considered as
principles, although there is no logical connection between them. The only
foundation for application of quantum principles is the fact that they
explain nonrelativistic quantum phenomena very well. They explain also
relativistic quantum phenomena, when they may be considered as small
correction to nonrelativistic phenomena. ACQP fails in explanation of
essentially relativistic quantum phenomena (for instance, pair production).
A possibility of modification of ACQP is connected mainly with application
of additional principles and suppositions, which change the structure of
ACQP. Possibility of dynamical modification (consideration of new dynamic
systems) is also take place, but in each special case one needs to use some
new ideas. ACQP in itself does not give foundation for such ideas, and this
makes the further development of ACQP to be difficult.

MCQP allows one to obtain new physical objects without any additional
suppositions. It is sufficient to remove or to vary some constraints,
imposed on the world function. The world function (\ref{a2.1}), determining
the microcosm structure is only the first rough approximation. If it is
necessary, the expression (\ref{a2.1}) can be modified in such a way, to
take into account influence of the matter distribution in the space-time
(curvature) and existence of new metric fields, generated by the possible
asymmetry of the world function \cite{R002a,R002b}.

Asymmetric world function describes the space-time, where the past and the
future are unequal geometrically. One cannot imagine such a thing in the
framework of Riemannian geometry. Expansion of the symmetric world function $%
\sigma \left( x,x^{\prime }\right) $ over powers of $\eta
^{i}=x^{i}-x^{\prime i}$ has the form 
\begin{equation}
\sigma \left( x,x^{\prime }\right) =\frac{1}{2}g_{ik}\left( x^{\prime
}\right) \eta ^{i}\eta ^{k}+\frac{1}{6}\sigma _{ikl}\left( x^{\prime
}\right) \eta ^{i}\eta ^{k}\eta ^{l}+...  \label{a4.8}
\end{equation}
where metric tensor $g_{ik}\left( x^{\prime }\right) $ describes the
gravitational field, and $\sigma _{ikl}\left( x^{\prime }\right) $ is
expressed via derivatives of metric tensor $g_{ik}\left( x^{\prime }\right) $%
. For asymmetric world function the same expansion has the form \cite
{R002a,R002b} 
\begin{equation}
\sigma \left( x,x^{\prime }\right) =\sigma _{i}\left( x^{\prime }\right)
\eta ^{i}+\frac{1}{2}\sigma _{ik}\left( x^{\prime }\right) \eta ^{i}\eta
^{k}+\frac{1}{6}\sigma _{ikl}\left( x^{\prime }\right) \eta ^{i}\eta
^{k}\eta ^{l}+...  \label{a4.9}
\end{equation}
where three coefficients $\sigma _{i}\left( x^{\prime }\right) $, $\sigma
_{ik}\left( x^{\prime }\right) $ and $\sigma _{ikl}\left( x^{\prime }\right) 
$ are independent, and each of them is connected with some metric
(geometric) field. Coefficient $\sigma _{i}\left( x^{\prime }\right) $
describes a ''vector field''\ which is strong and effective at small
space-time intervals. Coefficient $\sigma _{ik}\left( x^{\prime }\right) $
describes the second rank tensor field (gravitational field) which is strong
and effective at middle space-time intervals. Finally, $\sigma _{ikl}\left(
x^{\prime }\right) $ is connected with the third rank tensor field, which is
strong and effective at large space-time intervals. Maybe, this field is
connected with astrophysical problem of dark matter, when one fails to
explain observed motion of stars and galaxies by means of only gravitational
field.

At construction of MCQP \textit{one did not use any new hypotheses}. On the
contrary, flexibility and subtlety of MCQP methods are connected with remove
of unwarranted constraints and correction of mistakes in the approach to
geometry and to statistical description. In other words, MCQP satisfies the
Newton's criterion: ''Hypothesis non fingo.'' \textit{Only choice of true
space-time geometry is determined properties of physical phenomena in
microcosm}. This choice must be done in any case. But this choice may be
true, or not completely true. 

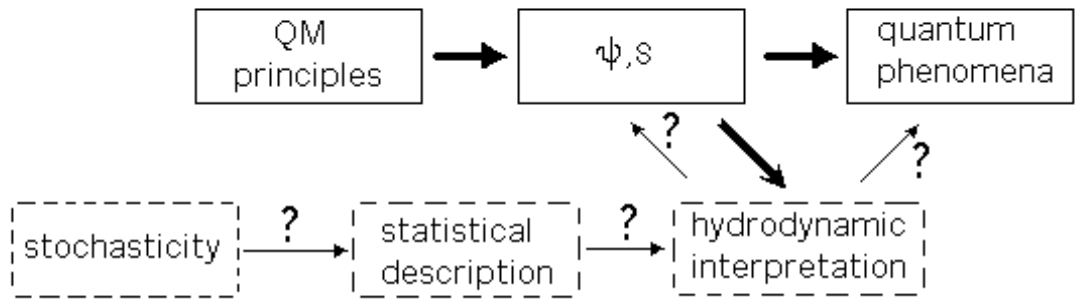
\end{document}